\documentclass[prl,twocolumn,showpacs,superscriptaddress]{revtex4}
\usepackage{amsfonts}
\usepackage{amsmath}
\usepackage{amssymb}
\usepackage{graphicx}

\newcommand{\be}{\begin{eqnarray}}
\newcommand{\ee}{\end{eqnarray}}

\begin{document}

\title{Liquid-vapor transition from a microscopic theory: Beyond the Maxwell construction}
\author{Alberto Parola}
\affiliation{Dipartimento di Fisica e Matematica, Universit\`a dell'Insubria, 
Via Valleggio 11, 22100 Como, Italy}
\author{Davide Pini}
\affiliation{Dipartimento di Fisica, Universit\`a degli Studi di Milano, 
Via Celoria 16, 20133 Milano, Italy}
\author{Luciano Reatto}
\affiliation{Dipartimento di Fisica, Universit\`a degli Studi di Milano, 
Via Celoria 16, 20133 Milano, Italy}



\begin{abstract}
A smooth cut-off formulation of the Hierarchical Reference Theory (HRT) is
developed and applied to a Yukawa fluid. 
The HRT equations are derived and numerically solved leading to:
the expected renormalization group structure in the critical region, non classical 
critical exponents and scaling laws, a convex free energy in the whole phase diagram
(including the two-phase region), 
finite compressibility at coexistence,
together with a fully satisfactory comparison with available numerical
simulations. This theory, which also guarantees the correct short range behavior of 
two body correlations, represents a major improvement over the existing liquid state theories.
\end{abstract}
\maketitle
Given that
the physics of the liquid-vapor phase transition is a textbook topic 
in thermodynamics, 
it is disappointing that so far no liquid-state theory has been 
able to describe it satisfactorily. In fact, when faced with phase coexistence,
mean-field approaches such as (generalized) van der Waals theories give 
a non-convex free energy, while integral equations \cite{caccamo} fail to converge altogether
in some domain inside the coexistence region. In both cases, the coexistence
boundary must be recovered via the Maxwell construction,
namely, by imposing thermodynamic equilibrium {\em a posteriori}. 
Aside of first-principles considerations, it should also be noted that this 
procedure may turn out to be cumbersome to implement, especially in the case
of integral equations, because of the need to circumvent the forbidden domain
and the ambiguities entailed by this procedure for theories lacking 
thermodynamic consistency. 
When dealing with mixtures of fluids, this becomes a serious hindrance to 
the theoretical determination of the phase diagram. 

The approach which comes closest to a realistic description 
of the liquid-vapor transition is the Hierarchical Reference Theory of fluids 
(HRT) \cite{hrt}, a genuine microscopic 
approach which implements the Renormalization Group (RG) method within
the liquid state framework. HRT is able to reproduce non classical critical 
exponents, scaling laws and, 
what is most relevant for this discussion, 
a rigorously convex free energy, so that   
flat isotherms at coexistence naturally emerge from the theory,
leading to accurate predictions of phase boundaries.
However, the implementation of HRT developed so far
is plagued by an unphysical divergence of the compressibility 
on the coexistence boundary \cite{first}. In liquid-state theory jargon, 
HRT forces the spinodal and the binodal curve to coincide. 
If flat isotherms are obtained at this price, one may have
some doubts about how reliable the description of the first-order phase 
transition given by HRT really is. 
This also prevents one from achieving any 
information on the possibile occurrence of metastable pure phases which, 
in the usual approaches, are located between the binodal and the spinodal.   

In this Letter we show that a novel implementation of HRT 
based on a smooth cut-off procedure \cite{smooth}, 
is able to reproduce correctly
the physics at the phase boundary and provide a criterion to discriminate 
between unstable and metastable states, while also improving 
on the representation of short range correlations.  
This qualifies HRT 
as the only liquid state theory able to provide a satisfactory
description of the liquid-vapor phase transition. 

Following the lesson of the RG \cite{rg}, 
HRT is based on the gradual introduction of 
(density) fluctuations starting from short wavelengths. 
The two body potential $v(r)$ is first split into the sum of a (repulsive) 
reference part 
$v_R(r)$ and a (mostly attractive) tail $w(r)$. 
The properties of the reference system, usually a hard sphere fluid,
are assumed to be known, 
and a sequence of intermediate systems is introduced, 
labeled by a parameter $t\in(0,\infty)$
``interpolating" between the reference and the physical model. 
The interaction potential of the $t$-system
is $v_t(r)=v_R(r)+w_t(r)$ where the Fourier components $\tilde w_t(k)$ of $w_t(r)$ 
are strongly suppressed for wave vectors $k \lesssim e^{-t}$. 
The change in the free energy of the system when the parameter $t$ 
undergoes an infinitesimal change can be evaluated exactly, leading to the HRT differential equation.

The purpose of the procedure outlined above is to suppress the liquid-vapor
transition throughout the whole sequence of intermediate $t$-systems, due to
the long range repulsive tail present in $w_t(r)$. Like in the RG,
the long-wavelength fluctuations which drive the 
phase transition must be allowed to develop only in the $t\to\infty$ limit,
when $w_t(r)$ tends to the physical attractive interaction $w(r)$. 
The sharp cut-off formulation of HRT, which has been successfully applied to
several model systems \cite{adv}, fulfills this condition. The sharp cut-off is defined by the 
choice $\tilde w_t(k)=\tilde w(k)$ for $k>e^{-t}$ while $\tilde w_t(k)=0$ for $k<e^{-t}$
and gives rise to an HRT equation which, close to the critical point and at long wavelengths, 
reproduces the Wegner-Houghton RG scheme \cite{wh}.
As previously discussed, this formulation suffers from some deficiency close to the first-order 
boundary, where a diverging compressibility is predicted by the sharp cut-off HRT. 
Remarkably, it has been recently shown, in the framework of scalar 
field theories, that this unphysical behavior can be eliminated by 
changing the cut-off procedure, i.e. by modifying the definition of 
the intermediate potentials $w_t(r)$ \cite{bonanno,phi4}.
If this property survives in the framework of a liquid state theory is 
not known and it is addressed here.
In this Letter, we adopt the following form: 
\begin{equation}
w_t(r)=w(r) - e^{-dt}\,\psi(t)\,w(r\,e^{-t})
\label{wt}
\end{equation}
where $d=3$ is the space dimensionality and $\psi(t)$ is a decreasing function of $t$ with $\psi(0)=1$ 
and asymptotic behavior $\psi(t) \propto e^{-2t}$ for $t\to\infty$. 
The precise definition of $\psi(t)$ will be discussed later. Note the $t$ dependence of the
range of the second (weakly repulsive) contribution to $w_t(r)$: in the large $t$ limit, the
amplitude of this term decreases while its range grows. 

The differential equation expressing the change in the free energy 
when the parameter $t$ (and hence the interaction $v_t$) is changed 
follows from first-order perturbation theory \cite{hansen}:
\begin{equation}
\frac{d A_t}{dt}=\frac{\rho^2}{2}\int d{\bf r} \,g_t(r) \,\frac{d\phi_t}{dt}
\label{hrt1}
\end{equation}
where $A_t$ is $(-\beta)$ times the Helmholtz free energy per unit volume of the fluid
interacting via $v_t(r)$ and $g_t(r)$ is the corresponding radial distribution function, 
while $\phi_t(r)=-\beta w_t(r)$. 
This equation is formally exact but, as usual in the HRT approach, 
it requires a closure relation expressing 
the two body correlations in terms of the free energy. Analogously to the standard implementation
of the sharp cut-off HRT \cite{adv}, we adopt the following Mean Spherical Approximation (MSA)-like 
representation \cite{hansen}:
\begin{eqnarray}
\label{closg}
g_t(r) &=& 0 \qquad\qquad\qquad\qquad 
{\rm for} \quad r<1 \\
c_t(r) &=& \phi_t(r) +\lambda_t \,\phi(r) \qquad 
{\rm for} \quad r>1
\label{closc}
\end{eqnarray}
where the direct correlation function $c_t(r)$ is related to $g_t(r)$ by the usual Ornstein-Zernike 
(OZ) equation \cite{hansen}. 
The key feature of this closure is the presence of the $\lambda_t$ parameter
which is introduced in order to enforce the compressibility sum rule:
\begin{equation}
\frac{\partial^2 A_t}{\partial\rho^2} = - \frac{1}{\rho} + \int d{\bf r} \,c_t(r)
\label{comp}
\end{equation}
Equations (\ref{hrt1}-\ref{comp}) form, together with the definition (\ref{wt}), 
a closed set of integro-differential equations for the thermodynamics and correlations of the 
model. The numerical solution of this problem poses a highly demanding computational tasks, which 
may be however simplified by choosing a particularly favorable model system: 
the Yukawa fluid, i.e a hard sphere fluid with attractive tail 
\begin{equation}
\phi(r)=-\beta w(r) = \frac{1}{T}\,\frac{e^{-z(r-1)}}{r}
\label{phi}
\end{equation}
where $T$ is the dimensionless temperature and $z$ is the inverse range. 
Lengths are normalized to the hard sphere diameter.
This one-parameter family of interaction potentials actually represents one of the most studied 
systems in liquid state theory: due to the simple analytical form of $w(r)$ and the flexibility 
due to the tunable inverse range parameter $z$, it provides a reasonable 
description of
simple fluids (for the celebrated choice $z=1.8$) as well as colloidal suspensions, where typically 
$z\gg 1$. A major advantage of this particular form of the interaction, follows from the
availability of an exact analytical solution of the OZ integral equation with the ansatz 
(\ref{closg},\ref{closc}) \cite{hoye}, 
which provides the explicit expression of the radial distribution function $g_t(r)$. 
By use of this solution, the right-hand side of the HRT equation (\ref{hrt1}) can be written 
in terms of $\lambda_t$ or, by use of Eq. (\ref{comp}), in terms of the free energy density $A_t$,
leading to a closed non linear partial differential equation. 

It is apparent the similarity between the Self Consistent Ornstein Zernike approximation 
(SCOZA) \cite{scoza} and this novel formulation of the HRT approach:
both theories satisfy Eqs. (\ref{closg},\ref{closc},\ref{comp}) and in both cases use is made of the 
analytic solution of the OZ equation for a Yukawa potential. Moreover, the consistency condition 
at the basis of SCOZA may be written in the form (\ref{hrt1}) for a specific choice of the 
turning-on procedure of the attractive interaction: in SCOZA the $t$ parameter (often identified with
the inverse temperature) only affects the amplitude of $w(r)$ without changing its range.
This seemingly minor difference has profound implications on the behavior of the theory 
in the critical region and close to the phase boundary. 

The cut-off function $\psi(t)$ in Eq. (\ref{wt}) has been chosen in order to guarantee 
the numerical stability of the HRT differential equation: 
$\psi(t)=(1+z\,t/2)^{-2}$ for $t<t^*$ and $\psi(t)=(\cosh\, t)^{-2}$
for $t>t^*$, where $t^*$ is defined by imposing the continuity of $\psi(t)$ \cite{note}. 
Full details will be given in a forthcoming publication. 

\begin{figure}
\includegraphics[height=4cm,width=8cm,angle=0]{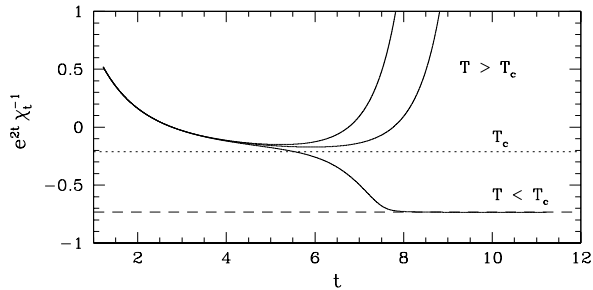}
\caption{Rescaled inverse compressibility 
for $z=1.8$
during the $t$-evolution at the 
critical density 
and reduced temperatures $(T-T_c)/T_c = 1.1 \times 10^{-5}, 2.6 \times 10^{-6}, -1.4 \times 10^{-5}$
from top to bottom. The dotted line is the fixed point 
critical value 
and the dashed line is the ``zero temperature" fixed point
obtained by a RG analysis of the HRT equation.}
\label{fix}
\end{figure}

\begin{figure}
\includegraphics[height=4cm,width=8cm,angle=0]{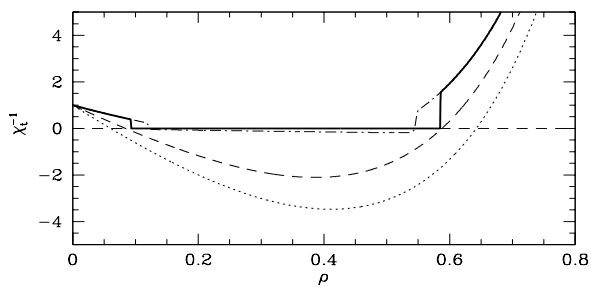}
\caption{Snapshots of the inverse compressibility along the density axis
for $z=1.8$ and $T=1.1$ at three different values of
the parameter $t$: $t=0$ (dotted line), $t=0.2$ (dashed line), $t=2.6$ (dot-dash line) 
and $t\to\infty$ (solid line).}
\label{disc}
\end{figure}

\begin{figure}
\includegraphics[height=4cm,width=8cm,angle=0]{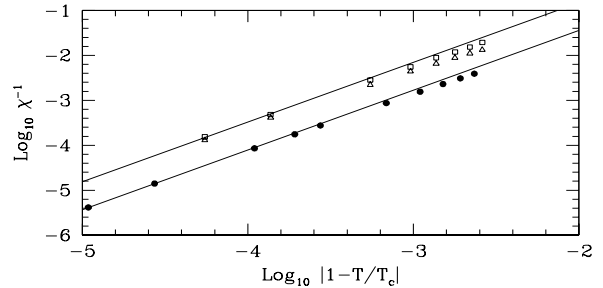}
\caption{Log-Log plot of the inverse compressibility as a function of the reduced temperature
above (circles) and below the critical temperature along the low-density (triangles) 
and the high-density (squares) branch of the binodal for $z=1.8$. 
Solid lines show the expected power law behavior defined by the exponent
of Table 1 ($\gamma=1.328$).}
\label{gamma}
\end{figure}

\begin{figure}
\includegraphics[height=6cm,width=8cm,angle=0]{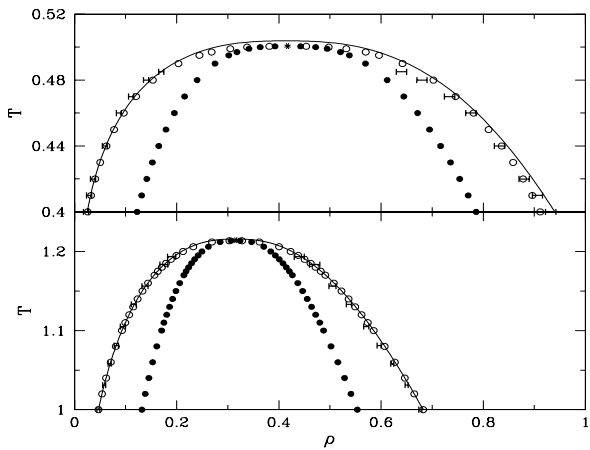}
\caption{Coexistence curve as obtained by HRT (open circles), SCOZA (solid line) and Monte Carlo 
(bars) for $z=5$ (upper panel) \cite{mex} and $z=1.8$ 
(lower panel) \cite{scoza}. The HRT spinodals are also shown 
(full circles).}
\label{phase}
\end{figure}

\begin{figure}
\includegraphics[height=7cm,width=8cm,angle=0]{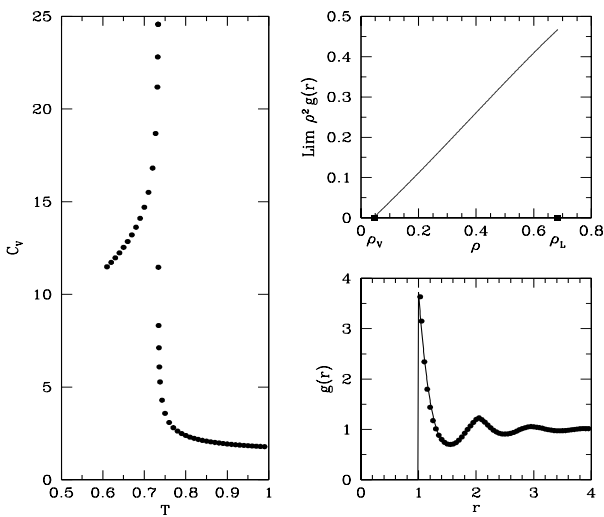}
\caption{
Left panel: HRT specific heat along the critical isochore $\rho=0.3605$ for $z=3$.
Lower-right panel: radial distribution function for 
$z=1.8$,
$T=1$, $\rho=0.8$. Line: HRT result; points: Monte Carlo simulations \cite{scoza}.
Upper-right panel: $r\to\infty$ limit of $\rho^2\,g(r)$ inside the coexistence curve
for $z=1.8$ and $T=1$.} 
\label{cvgr}
\end{figure}

Close to the critical point and in the $t\to\infty$ limit, the HRT equation (\ref{hrt1})
simplifies and, when the thermodynamical variables are properly rescaled, acquires a RG structure.
The precise form of the rescaled equation actually depends on the specific form of the 
attractive part of the interaction $\phi(r)$ (\ref{phi}), which in HRT plays the role of the 
smooth cut off of the RG approach. A fixed point analysis, similar to that performed for 
a $\Phi^4$ theory \cite{phi4}, shows that the HRT equation satisfies scaling and hyperscaling 
with the non-classical critical exponents shown in Table 1. 
\begin{table}
\vskip 0.2cm
\begin{tabular}{|c||c|c|c|c|c|c|}
\hline
Exponent & $\alpha$   & $\beta$   & $\gamma$   & $\delta$    & $\eta$ & $U_2=C_+/C_-$ \\ 
\hline
``Exact"    & 0.110 & 0.327 & 1.237 &   4.789 & 0.036 & 4.76  \\
\hline
HRT & 0.01 & 0.332 & 1.328 & 5 & 0 &  4.16 \\
\hline
\end{tabular}
\caption{HRT estimates of the critical exponents and compressibility
amplitude ratio in three dimensions for $z=1.8$
compared to the exact values \cite{field}.}
\end{table}
The numerical solution of the 
full HRT equation (\ref{hrt1}) allows to justify the fixed point analysis on microscopic 
grounds: very close to the critical point, the 
quantity 
\begin{equation}
\chi^{-1}_t = - \rho \,\left [\frac{\partial^2 A_t}{\partial\rho^2} +\psi(t)\int d{\bf r} \phi(r) \right ]
\label{chi}
\end{equation}
when multiplied by the rescaling factor $e^{2t}$, 
falls into the basin of attraction of the fixed point, as shown 
in Fig. \ref{fix}.  
However, at very long wavelength (i.e. for $t\to\infty$), 
the small deviations from the critical point drive the system either to the ``infinite temperature"
or to the ``zero temperature" fixed point. 
The physical meaning of $\chi^{-1}_t$ is apparent from its definition:
it is proportional to the inverse compressibility (\ref{comp}) supplemented by the mean field
contribution associated to the residual part of the potential $w(r)-w_t(r)$. 
In order to understand how the singularity associated to the first-order liquid-vapor transition 
develops within HRT, it is instructive to follow the ``evolution" of the inverse compressibility $\chi_t^{-1}$
for a fixed temperature below the critical point and different values of $t$ (see Fig. \ref{disc}). 
Long wavelength fluctuations force the inverse compressibility to vanish identically 
in the $t\to\infty$ limit inside the binodal, as customary in the HRT approach \cite{first}. The novel
feature displayed by the smooth cut-off formulation, and clearly visible in 
Fig. \ref{disc}, is the jump of $\chi_\infty^{-1}$ across the phase boundary. 
A closer inspection of the $t$-evolution of $\chi^{-1}$ at coexistence reveals that 
the approach toward zero proceeds differently deep inside the binodal, where $\chi^{-1}$
remains negative throughout the evolution, and close to the phase boundary, where long wavelength 
fluctuations first drive the system towards stability ($\chi^{-1} > 0$) and then push the 
inverse compressibility to zero. It is tempting to identify the boundary between these 
two regimes as the boundary of metastability, i.e. the spinodal curve. 
A power law fit of the compressibility in the critical region $\chi\sim C\,|T-T_c|^{-\gamma}$, 
shown in Fig. \ref{gamma}, is fully consistent with the critical exponent $\gamma$  
reported in Table 1 both above and below the critical temperature.
The amplitude ratio $C_+/C_-$, also shown in Table 1, agrees well with the field theoretical 
expectation. 
The phase diagram of a Yukawa fluid with $z=1.8$ and $z=5$ is compared with Monte Carlo 
simulations and SCOZA results in Fig. \ref{phase}. 
The constant volume specific heat can be obtained by differentiation of the free energy density at convergence
$A_\infty(\rho,T)$. In Fig. \ref{cvgr} we plot the specific heat per particle $C_v$ along the
critical isochore: the data below the critical temperature are obtained by use of the free
energy inside the coexistence curve, which is directly accessible in HRT.   
A typical radial distribution function of a $z=1.8$ Yukawa liquid as obtained by HRT is shown in Fig. \ref{cvgr}.
The core condition is exactly fulfilled due to Eq. (\ref{closg}) and the
use of the analytical solution of the OZ equation. However, the linear dependence 
of $c(r)$ on the attractive interaction $w(r)$ 
implied by the MSA-like closure (\ref{closc}) limits the accuracy of HRT for the
description of the local structure of the fluid, particularly at low density. 
Nevertheless,
the accurate treatment of the physics 
underlying the first-order transition has important consequences on the form of $g(r)$ 
inside the binodal.
In the two-phase region, density correlations are linear combinations of those of the two stable phases (liquid $L$
and vapor $V$). This implies the exact relation
$\lim_{r\to\infty}\rho^2 g(r)\,= \rho(\rho_V+\rho_L)-\rho_V\rho_L$
which is satisfied by the numerical solution of the smooth cut-off HRT 
equation, as displayed in Fig. \ref{cvgr}.

In summary, we have shown how a smooth cut off formulation of HRT provides a 
consistent picture of the equilibrium thermodynamics of a fluid, including the
complex, singular behavior at first- and second-order phase transitions, 
a distinction between unstable and metastable states, and 
quantitative predictions for the coexistence curve, equation of 
state and specific heat. 
The present formulation of the theory is specific to a single Yukawa interaction, but the available
analytical solution of the OZ equation for a sum of an arbitrary number of Yukawa
potentials foreshadows possible generalizations. A 
better representation of 
correlations may be also achieved within the HRT framework, either by adopting parametrizations
more elaborate than Eqs. (\ref{closg},\ref{closc}), or by closing the hierarchy at the level 
of the second equation, which embodies the effects of density fluctuations on the structure of the
fluid \cite{adv}. 
Applications to binary mixtures \cite{mix} and non-uniform fluids \cite{orlandi}
are also possible.

We thank N.B. Wilding for providing the Monte Carlo data and we 
acknowledge support from EC, contract number MRTN-CT2003-504712.


\begin{thebibliography}{10}
\bibitem{note} 
We verified that the results do not differ appreciably when alternative 
choices of the function $\psi(t)$ are adopted.
\end{thebibliography}
\end{document}